\documentclass[12pt,aps,pdftex]{revtex4}
\usepackage{amsmath,amssymb,url,graphics,graphicx}
\usepackage[font=scriptsize,justification=RaggedRight]{caption}
\pdfoutput=1
\begin{document}
\title{Networks of Economic Market Interdependence \\and Systemic Risk}
\author{Dion Harmon}
\affiliation{New England Complex Systems Institute, Cambridge, MA 02142, USA}
\author{Blake Stacey}
\affiliation{New England Complex Systems Institute, Cambridge, MA 02142, USA}
\author{Yavni Bar-Yam}
\affiliation{New England Complex Systems Institute, Cambridge, MA 02142, USA}
\author{Yaneer Bar-Yam}
\email[To whom correspondence should be addressed: ]{yaneer@necsi.edu}
\affiliation{New England Complex Systems Institute, Cambridge, MA 02142, USA}
\date{March 6, 2009; revised November 11, 2010}

\begin{abstract}
The dynamic network of relationships among corporations underlies
cascading economic failures including the current economic crisis, and
can be inferred from correlations in market value fluctuations. We
analyze the time dependence of the network of correlations to reveal
the changing relationships among the financial, technology, and basic
materials sectors with rising and falling markets and resource
constraints. The financial sector links otherwise weakly coupled
economic sectors, particularly during economic declines. Such links
increase economic risk and the extent of cascading failures. Our
results suggest that firewalls between financial services for
different sectors would reduce systemic risk without hampering
economic growth.
\end{abstract}

\maketitle

The global economy is a highly complex system \cite{ref:dcs} whose
dynamics reflects the connections among its multiple components, as
found in other networked
systems \cite{ref:barrat2008,ref:christakis2008,ref:hidalgo2007}. A
common property of complex systems is the risk of cascading failures,
where a failure of one node causes similar failures in linked nodes
that propagate throughout the system, creating large scale collective
failures. Economic risks associated with cascading financial losses
are manifest in the current economic crisis \cite{ref:greenlaw2008}
and the earlier Asian economic crisis \cite{ref:radelet1998}, but are
not considered in conventional measures of investment risk
\cite{ref:jorion2006}.

A central question is the role that complex systems science can play
in informing regulatory policy that preserves the ability of markets
to promote economic growth through freedom of investment, while
protecting the public interest by preventing financial meltdowns due
to systemic risk.

Characterizing the network of economic dependencies and its
relationship to risk is key
\cite{ref:mantegna1999,ref:garas2008,ref:schweitzer2009,ref:smith2009,ref:emmert-streib2010}. The
dependencies among organizations involve large numbers of factors,
including competition for capital and labor, supply and demand
relationships among organizations that deliver common end products or
rely upon common inputs, natural disasters and climate conditions,
acts of war and peace, changes of government or its policies including
economic policy such as interest rates, and geographic
association. Quantifying such dependencies, {\em e.g.,} through
Leontief models \cite{ref:carvalho2008,ref:leontief1986}, is difficult
because many of the dependencies are non-linear and driven by
socio-economic events not included in these models. Also, behavioral
economics \cite{ref:barberis2003,ref:delong1990,ref:delong1990b}
suggests that under some conditions collective investor behavior, {\em
  e.g.,} from perceptions of value, may have significant
effects. Reflecting both fundamental and behavioral interactions,
correlations in market value of firms can serve as a measure of the
perceived aggregate financial dependence and quantify ``herding''
behavior in collective fluctuations. Moreover, price correlations are
directly relevant to measures of risk.

\begin{figure}[h]
\includegraphics[width=18cm]{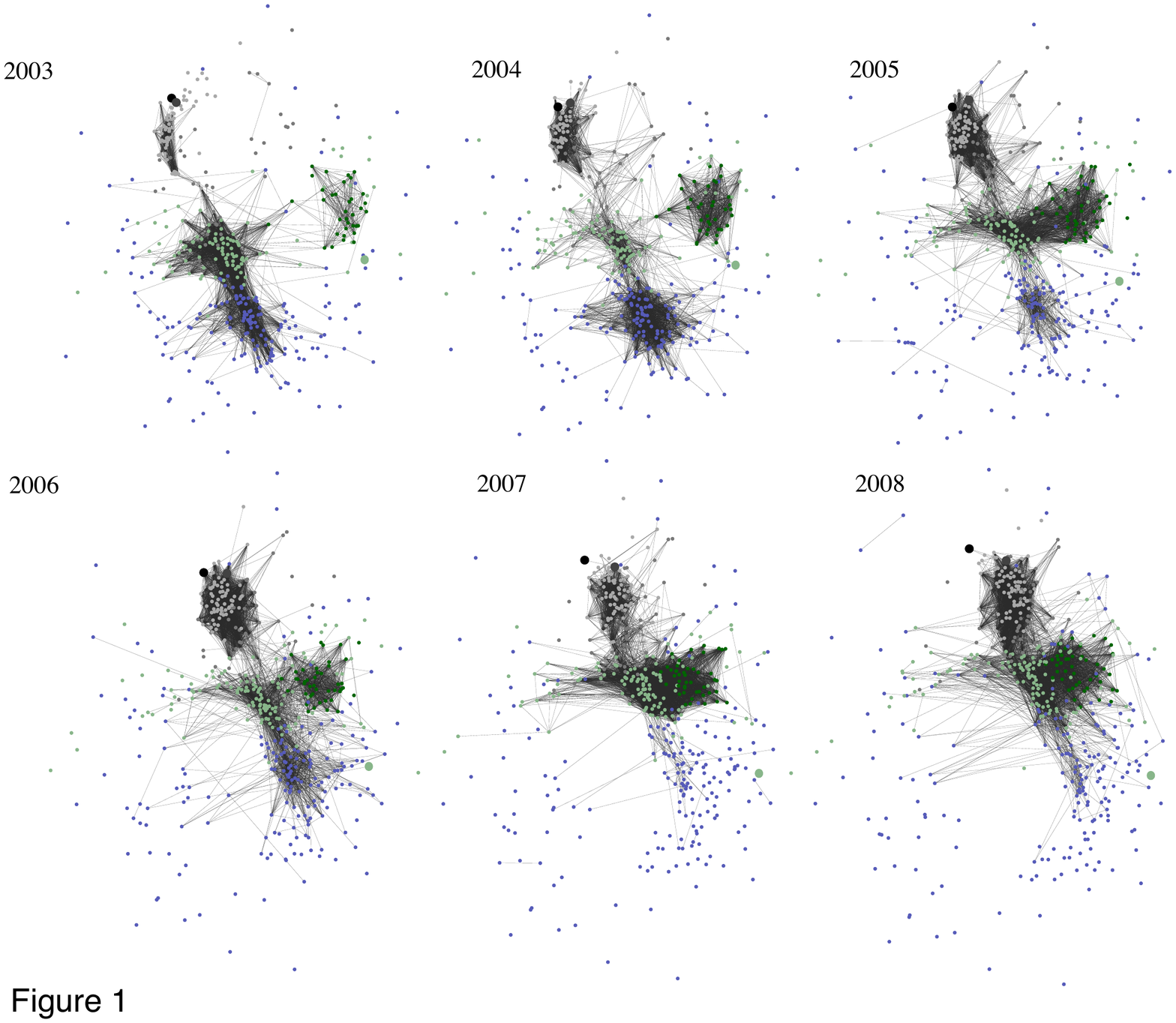}
\caption{\label{fig:1} Network of correlations of market daily
    returns for years as indicated. Dots represent individual
    corporations colored according to economic sector: technology
    (blue), basic materials including oil companies (light grey) and
    others (dark grey), and finance including real-estate (dark green)
    and other (light green). Links shown are the highest 6.25\% of
    Pearson correlations of $\log(p(t)/p(t-1))$ time series, where
    $p(t)$ are adjusted daily closing prices of firms
    \cite{ref:yahoo}, in each year. Larger dots are spot oil prices at
    Brent, UK and Cushing, OK (black) and the price of ten year
    treasury bonds (green). }
\end{figure}

We constructed a network of dependencies among 500 corporations having
the largest stock trading volume, augmented with several economic
indices (oil prices, and bond prices reflecting interest rates). We
formed a network where links are present for the highest correlations
in daily returns in each year from 2003 to 2008.  In order to display
the effect of changes over time, we constructed a single network over
all years, with each corporation in a particular year represented by a
node linked to itself in the previous and next year. Each year is
separately shown in Figure~\ref{fig:1}. We included only economic
sectors that are significantly self-correlated, as the larger network
constructed from the entire market obscures key insights. Previous
correlational analyses have described how correlations may arise from
external forces across the market (arbitrage pricing theory
\cite{ref:chamberlain1983,ref:ross1976}) or used correlations to
characterize sectors and market crashes (econophysics
\cite{ref:mantegna2000,ref:onnela2003}). This work lacks an
understanding of the economic origins of changes in dependencies and
their policy implications. We examine variations of within- and
between-sector correlations, arising from non-linear effects, for
information about changes in economic conditions prior to and during
the economic crisis.

The study of network community properties often requires careful
analysis \cite{ref:fortunato2010}.  In our case, the observations we
describe are manifest visually and were also tested statistically.  In
particular, apparent trends were tested using the $t$-statistic of
differences in link densities within and between sectors (merging), or
the minimum of this statistic between one sector and each of the
others (self-clustering). Sectors are statistically linked (unlinked)
to an index, if the $t$-statistic comparing links to the index
relative to the link density of the graph is above 4 (below 2).

The following observations and trends from 2003 through 2008 are
apparent and quantifiable: In 2003 there is a separate cluster of real
estate related financial institutions (dark green), which over time
merges into the larger financial cluster (green) (not merged through
2004 quarter 4, $p<10^{-10}$, from 2007 quarter 2 to 2008 quarter 3,
$p \geq 0.18$.). The technology sector (blue), while strongly
clustered during economic growth (2003-2006), becomes relatively
weakly clustered during the economic crisis (2008) (self-clustering
statistic has negative slope, $p<10^{-66}$, and changes sign in 2008,
$p<10^{-10}$). Interest rates (larger green dot) are sometimes related
to the technology cluster (linked for 8 out of 26 quarters). The oil
sector (grey) is highly clustered (any other sector is separate,
$p<10^{-13}$), and over time becomes increasingly linked to the rest
of the basic materials cluster (dark grey) (positive slope,
$p<10^{-45}$), which itself becomes more connected to the technology
cluster (positive slope, $p<10^{-64}$).  The oil cluster is only
sometimes correlated to oil prices (large black dots) (linked for 7 of
27 quarters). We will show that the network dynamics are consistent
with the sequence of economic events of the financial crisis
\cite{ref:greenlaw2008}. In traditional external factor models and
models of collective behavior in interacting systems \cite{ref:dcs},
correlations are constant over time, but recent models have introduced
the fitting of dynamical correlations of market indices
\cite{ref:cappiello2007,ref:engle2002}. We will show that changes in
correlations among corporations can be understood using intuitive
models for this period of time.

\begin{figure}[p]
\includegraphics[width=18cm]{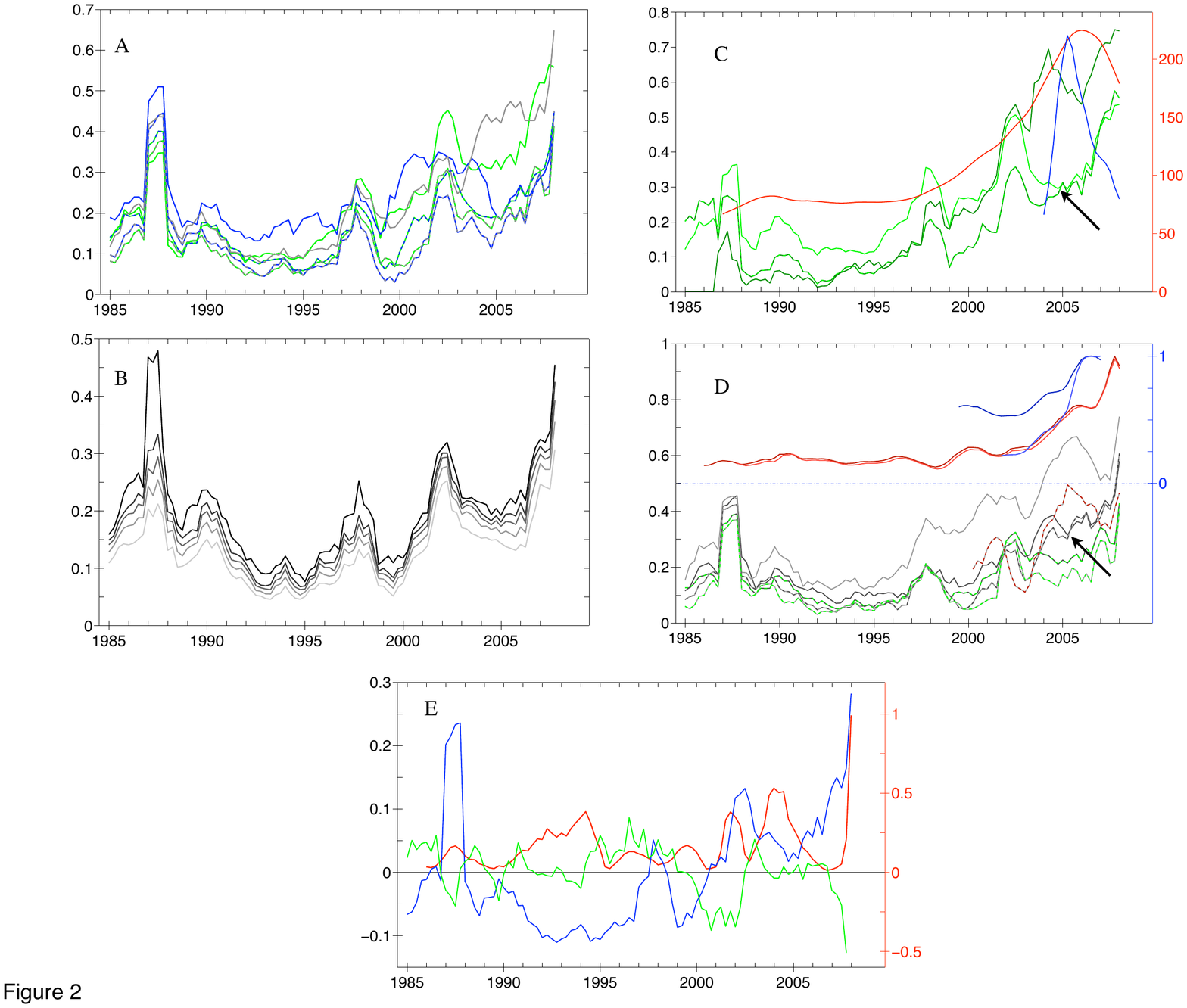}
\caption{\label{fig:2} Market correlations and external events from
  1985 to 2008. {\bf A:} The average strength of correlations within
  and between economic sectors. Sectors included are finance (green),
  technology (blue), and basic materials (grey), double colored lines
  are correlations between sectors (blue-grey, blue-green, and
  grey-green). Correlations are calculated using twelve month windows,
  shifted quarterly from Jan 1985 through Jan 2008. {\bf B:} Average
  correlations among stocks from all economic sectors. Black to light
  grey lines omit in each 12 month period the highest 0, 2, 5, 10, 20
  absolute average return days respectively. {\bf C:} Financial sector
  correlations separated into real estate related (dark green), other
  (light green), and between these sectors (hatched light and dark
  green) using left axis scale. The arrow indicates the effective
  merger of the sectors. Also shown are a housing price indicator
  (red, using right axis scale) \cite{ref:sp} and the search volume on
  Google for ``housing bubble'' (blue, arbitrary units)
  \cite{ref:google}. {\bf D:} Basic materials sector correlations
  separated into oil (light grey) and others (dark grey) as well as
  finance (green), with mixed color lines reflecting inter-sector
  correlations (left axis scale). Also shown are prices of spot oil in
  Brent, UK (light red) and Cushing, OK (dark red), aluminum (light
  blue) and copper (dark blue) normalized to maximum values (right
  axis, both oil prices are normalized to the maximum of Cushing,
  OK). Average correlation of oil price in OK with the oil sector is
  shown (red/grey hatched). Arrow is the merger of oil and other basic
  materials. {\bf E:} Rolling average correlations of the sectors in A
  (blue, left axis scale) shown with market value change (green, left
  axis scale, the return of S\&P500 index). Market declines (negative
  returns) coincide with higher than average market correlations
  ($p<0.02$). Also shown are effective limits on interbank loans (red,
  right axis scale, the difference between the London Interbank
  Offered Rate (LIBOR) and the Federal Funds Overnight Rate
  (annualized) at the beginning of each quarter, divided by the
  latter), having high values in the current economic crisis.}
\end{figure}

Specific external events can be identified whose timing coincides with
observed changes in correlations. Fig.~\ref{fig:2}C shows that the
merger of the real estate and other financial sectors stocks coincides
with both a peak in search frequency for ``housing bubble'' on Google
\cite{ref:choi2009}, and a turning point in the behavior of housing
prices ($p<10^{-3}$). This timing is consistent with the understanding
\cite{ref:greenlaw2008} that the decline in housing prices triggered
the financial system crisis due to large investments in mortgage
backed securities across the financial sector. Fig.~\ref{fig:2}D shows
a potential role of critical resources: first, in the changing
coupling of the basic materials sector to other parts of the economy;
second, in the changing coupling of oil sector to oil prices, which is
only one of the factors affecting the oil industry. Nonlinearity due
to dramatic increases in prices can readily be explained because they
are additive components of fundamental economic factors, {\em i.e.,}
costs of production. When commodity prices are low, other components
dominate, but when commodity prices are high they have larger effects,
so the fractional variation is nonlinearly related to the total. The
proximate coincidence of the severe commodities price increases
\cite{ref:bernanke2008} with the housing crisis ($p<10^{-5}$) may be
understood either through fears of commodity shortages due to rapid
growth, or the transfer of investment from the housing sector to
commodities \cite{ref:cabellero2008}---investment demand rather than a
use demand surge.  This is consistent with the observation that
economic growth by itself does not cause high correlations. However,
general considerations of the role of constraints imply that when
growth encounters the limit of available resources, increased
correlations should occur as changes in one sector impact resource
availability for another. Note that the correlations are primarily
positive---commodity values rise with increasing financial sector
values---consistent with fears of growth causing shortages or
increasing investment demand. Negative correlations would be expected
if commodity prices actually constrained economic growth.

Limiting investments ({\em i.e.,} limiting capital-to-asset ratios) in
order to moderate risk directly influences opportunities for
growth. However, our results also point to a different strategy, which
recognizes that financial institutions cross-link otherwise weakly
correlated economic sectors. The key is that economic couplings among
companies propagate the effect of failures. If economic entity G fails
in a financial obligation to entity H, the impact on H may affect
other entities J and K, that are linked to H, even if their activity
has nothing to do with G. Conversely, while a small capital-to-asset
ratio may be risky for a particular institution, if the investments
are within a particular economic sector the failure of that
institution is unlikely to cause economy wide repercussions. Thus,
segregating financial relationships, particularly among activities
that are not otherwise related, or are weakly related, reduces
systemic risk.

The idea that separations between components of the financial sector
contribute to economic stability was a key aspect of legislation to
stabilize the American banking system after the market crash of
1929. The Glass--Steagall Act of 1933 \cite{ref:fdic2007,ref:heakal}
separated investment banking from consumer (retail) banking to prevent
the fluctuations from other parts of the economy affecting consumer
banking. This Act was progressively eroded until its repeal in 1999
\cite{ref:gramm1999}. Other historical forms of separation imposed by
law or by practice included the separation of savings and loan
associations and insurance providers from commercial and investment
banking, as well as geographic separation by state
\cite{ref:fdic2007,ref:gramm1999}. While many effects contribute to
correlations in economic activity
\cite{ref:carvalho2008,ref:horbath1993}, nonlinearities associated
with investment during market declines support the historical
intuition that regulating these dependencies is more critical than
regulating those arising from, {\em e.g.,} supply chains. One of the
arguments in favor of deregulation was that banks, by investing in
diverse sectors, would have greater stability \cite{ref:heakal}. Our
analysis implies that the investment across economic sectors itself
creates increased cross-linking of otherwise much more weakly coupled
parts of the economy, causing dependencies that increase, rather than
decrease, risk. Quite generally, separation prevents failure
propagation and connections increase risks of global
crises. Subdivision is a universal property of complex systems
\cite{ref:dcs,ref:simon1997}. An increase in separation of financial
services is likely to entail costs, and the cost-benefit tradeoffs of
imposing particular types of separation are yet to be determined.

In summary, complex systems science focuses on the role of
interdependence, a key aspect of the dynamical behavior of economic
crises as well as the evaluation of risks in both ``normal'' and rare
conditions. We have analyzed the dynamics of correlational
dependencies in rising and falling markets. The impact on the economic
system of repeals of Depression-era government policies is becoming
increasingly manifest through scientific analysis of the current
economic crisis. Previous studies \cite{ref:pozen2008} showed that
repeal of the ``uptick rule'' in 2007 reduced economic stability by
reducing returns and increasing fluctuations of the securities
market. This study suggests that erosion of the Glass--Steagall Act,
the consolidation of banking functions, and cross sector investments
eliminated ``firewalls'' that could have prevented the housing sector
decline from triggering a wider financial and economic crisis.

Acknowledgements: We thank James H. Stock, Jeffrey C. Fuhrer and
Richard Cooper for helpful comments.

\end{document}